\documentclass[a4paper,12pt]{article}
\usepackage{fullpage}
\usepackage{multicol}
\usepackage{multirow}
\usepackage{hyperref}

\newcommand{\EuroControl}{EUROCONTROL}

\begin{document}

\title{{\bf Automatic Airspace Sectorisation: \\
    A Survey \thanks{This work has been co-financed by the European
      Organisation for the Safety of Air Navigation (\EuroControl)
      under its Research Grants programme (contract 08-1214447-C).
      The content of the work does not necessarily reflect the
      official position of \EuroControl\ on the matter.}
  }}

\author{Pierre Flener and Justin Pearson \\
  Department of Information Technology \\
  Uppsala University, Box 337, SE -- 751 05 Uppsala, Sweden \\
  \url{Firstname.Surname@it.uu.se}}

\date{Revision of \today \\ (first version on November 30, 2012)}

\maketitle

\begin{abstract} \noindent
  Airspace sectorisation provides a partition of a given airspace into
  sectors, subject to geometric constraints and workload constraints,
  so that some cost metric is minimised.  We survey the algorithmic
  aspects of methods for automatic airspace sectorisation, for an
  intended readership of experts on air traffic management.
\end{abstract}

\section{Introduction}

Airspace \emph{sectorisation} provides a partition of a given airspace
into a given (or upper-bounded, or minimal) number of control sectors,
subject to geometric constraints and workload constraints, so that
some cost metric is minimised.  We parametrise the concept in
Section~\ref{sect:dim}.

Airspace sectorisation is not to be mixed up with airspace
\emph{configuration}, which provides a schedule for the grouping and
splitting of elementary sectors into control sectors that are suitable
for a given number of available controllers and the expected traffic
structure.  This survey does not cover papers showing how to compute
such \emph{sector opening schemes}, such as the works
of~\cite{Barbosa:CASPT01,Verlhac:EEC05}, even though there is an
overlap between both problems.  Configuration is by definition a
(pre-)tactical action, whereas sectorisation is either strategic or
(pre-)tactical, depending on the inputs, so that a sectorisation model
can in principle be re-used within a configuration model.  However, a
configuration model also has time variables for scheduling purposes
and aims to minimise the total delay over a given time interval, but
these temporal aspects are absent from sectorisation.  Further, a
configuration model includes a transition cost incurred at every
switch between configurations, but static sectorisation does not
consider such a cost.  Hence a configuration model is much harder to
re-use properly for a sectorisation model, as projecting away time
variables and re-configuration costs most likely enables a better
model.

This paper is a survey of automatic airspace sectorisation methods.
To the best of our knowledge, there is only one prior survey, namely
\cite{Zelinski:DASC11}, which actually replaces two prior surveys by
the first author of that survey.  To distinguish our survey from that
survey, our survey has the following caveats:
\begin{itemize}
\item Our survey is intended for experts on air traffic control (ATC)
  and air traffic management (ATM) and is thus not self-contained:
  technical jargon (in use at \EuroControl\ for instance) as well as
  the rationales behind existing procedures and investigated research
  topics are assumed to be known by the reader.  For definitions of
  concepts such as elementary sector, control sector, air traffic
  control centre (ATCC) = area control centre (ACC), functional
  airspace block (FAB), ATC functional block (AFB), sector capacity,
  sector load, etc, see~\cite{ASTRA:KER:survey} for instance.
\item Our survey is written by experts on computing science: no
  evaluation of the realism of the combination of inputs, parameters,
  outputs, and experiments of automatic airspace sectorisation tools
  is made, as that is the realm of ATM experts.  See
  \cite{Zelinski:DASC11} for a comparison of six of the methods
  discussed in our survey.
\item Our survey is only about the algorithmic aspects of airspace
  sectorisation tools: for instance, we discuss no papers giving only
  rationales for the constraints and objective functions that can be
  used in sectorisation.
\end{itemize}
We do not claim that our survey is complete.  Indeed, the literature
is growing rapidly nowadays.  However, many papers are archived at
pay-sites that we (as academic computer scientists from a university
without a transportation research centre) have no free access to, so
it takes some effort to get hold of these papers.  Also, some
conferences seem to accept papers that are almost devoid of the
technical details that would in principle allow the reader to
reproduce the results, so it is hard to compare such papers to the
others.  Whenever an author (team) has published multiple papers, we
examine the most recent publication (that we could find).

The rest of this survey is organised as follows.  In
Section~\ref{sect:dim}, we introduce the classification criteria used
in the survey proper, which is in Section~\ref{sect:survey}.  In
Section~\ref{sect:concl}, we conclude.

\section{Airspace Sectorisation: Classification Criteria}
\label{sect:dim}

Our survey in Section~\ref{sect:survey} of the literature on airspace
sectorisation is structured around the following classification
criteria.

\paragraph{Approach.}
We distinguish between two approaches to airspace sectorisation.
\begin{itemize}
\item In a \emph{graph-based} model, a graph is constructed whose
  vertices represent the intersections of the existing trajectories,
  and whose edges thus represent segments of the existing
  trajectories.  The core problem of sectorisation is then essentially
  the NP-complete combinatorial problem of graph
  partitioning~\cite{GareyJohnson:NP}.  A graph partition does not
  define the sector boundaries, so actual sectors have to be
  constructed from the resulting vertex sets in a geometric
  post-processing step.
\item In a \emph{region-based} model, the airspace is initially
  partitioned into some kind of regions that are smaller than the
  targeted sectors, so that the combinatorial problem of partitioning
  these regions in principle needs no geometric post-processing step.
\end{itemize}

\paragraph{Frequency.}
Airspace sectorisation can be invoked with different frequencies:
\begin{itemize}
\item \emph{Static}: Sectorisation is strategic or pre-tactical.
\item \emph{Dynamic}: Sectorisation is tactical, but occurs at
  pre-determined times (so as to be different from configuration).
\end{itemize}

\paragraph{Input Granularity.}
Region-based models can start from regions of (any combination of)
different granularities:
\begin{itemize}
\item A mesh of blocks of the same size and shape.
\item ATC functional blocks (AFBs).
\item Elementary sectors, namely the ones of the existing
  sectorisation.  This leads to a further
  distinction~\cite{Zelinski:DASC11}:
  \begin{itemize}
  \item \emph{Base-line}: The output sectorisation should be
    reasonably close to the input one, usually because the
    sectorisation is dynamic and bears a transition cost.
  \item \emph{Free-form}: The output sectorisation can be arbitrarily
    far from the input one.
  \end{itemize}
\item Control sectors.
\item Area of specialisation (AOS).
\item Air traffic control centre (ATCC).
\end{itemize}

\paragraph{Output Granularity.}  
An airspace sectorisation can be computed at different levels of
granularity:
\begin{itemize}
\item Elementary sectors.
\item Control sectors.
\item Functional airspace blocks (FABs).
\item Area of specialisation (AOS).
\item Air traffic control centre (ATCC).
\end{itemize}

\paragraph{Dimensionality.}
An airspace sectorisation can be computed in different (numbers of)
dimensions:
\begin{itemize}
\item 2D: The sectorisation is only defined (and tested) in two
  dimensions (longitude and latitude), with the generalisation to 3D
  being considered to be straightforward.
\item 2.5D: The sectorisation is only computed in two dimensions,
  because layers of the 3D airspace are considered to be independent.
\item 3rd D: The sectorisation preserves the 2D boundaries of the
  input regions but can readjust them in the third dimension
  (altitude).
\item 3D: The sectorisation is computed in three dimensions.
\end{itemize}

\paragraph{Constraints.}
Airspace sectorisation aims at satisfying some constraints.  The
following constraints have been found in the literature, so that a
subset thereof is chosen for a given tool:
\begin{itemize}
\item \emph{Balanced workload}: The workload of each sector must be
  within some given imbalance factor of the average across all
  sectors.
\item \emph{Bounded workload}: The workload of each sector must be
  below some upper bound.
\item \emph{Balanced size}: The size of each sector must be within
  some given imbalance factor of the average across all sectors.
\item \emph{Minimum dwell time}: Every flight entering a resulting
  sector must stay within it for a given minimum amount of time (say
  two minutes), so that the coordination work pays off and that
  conflict management is possible.
\item \emph{Minimum distance}: Each existing trajectory must be inside
  each resulting sector by a minimum distance (say ten nautical
  miles), so that conflict management is entirely local to sectors.
\item \emph{Convexity} of the resulting sectors.  Convexity can be in
  the usual \emph{geometric} sense, or \emph{trajectory-based} (no
  flight enters the same sector more than once), or more complex.
\item \emph{Connectedness}: A sector must be a contiguous portion of
  airspace and can thus not be fragmented into a union of unconnected
  portions of airspace.
\item \emph{Compactness}: A sector must have a geometric shape that is
  easy to keep in mind.
\item \emph{Non-jagged boundaries}: A sector must have a boundary that
  is not too jagged.
\end{itemize}
For each sector, there are three kinds of workload: the
\emph{monitoring workload}, the \emph{conflict workload}, and the
\emph{coordination workload}; the first two workloads occur inside the
sector, and the third one between the sector and an adjacent sector.
The quantitative definition of workload varies strongly between the
surveyed papers.  In the following, the word ``workload'' refers to
the sum of all three workload terms, but we will specify the actually
used workload terms whenever they can be inferred from a paper.

\paragraph{Constraint Types.}
Airspace sectorisation constraints can be of different types:
\begin{itemize}
\item A \emph{hard} constraint must be satisfied, whereas a
  \emph{soft} constraint can be violated, although its satisfaction
  earns a numeric bonus.
\item An \emph{explicit} constraint is part of the model, whereas an
  \emph{implicit} constraint is enforced by side effect, either
  because it is logically \emph{implied} by the explicit constraints
  or because its satisfaction is an \emph{invariant} of the solution
  process.
\end{itemize}

\paragraph{Cost Function.}
Airspace sectorisation aims at minimising some cost.  The following
costs have been found in the literature, so that a subset thereof is
combined into the cost function for a given tool, the subset being
empty if sectorisation is not seen as an optimisation problem:
\begin{itemize}
\item \emph{Coordination cost}: The cost of the total coordination
  workload between the resulting sectors must be minimised.
\item \emph{Transition cost}: The cost of switching from the old
  sectorisation to the new one must be minimised.
\item \emph{Workload imbalance}: The imbalance between the workload of
  the resulting sectors must be minimised.
\item \emph{Number of sectors}: The number of sectors must be
  minimised.
\item \emph{Entry points}: The total number of entry points into the
  resulting sectors must be minimised.
\item If any of the constraints above is soft, then there is the
  additional cost of minimising the number of violations of soft
  constraints.
\end{itemize}
A workload cost results from applying a function to the workload to
which it pertains; if this is the identity function, then we talk of
coordination workload rather than coordination cost.

\paragraph{Technology.}
An airspace sectorisation model can be implemented using (any
combination of) different algorithm design methodologies or
optimisation technologies:
\begin{itemize}
\item Stochastic local search (SLS), see \cite{Hoos:SLS} for instance.
\item Constraint programming (CP), see \cite{constraintshandbook} for
  instance.
\item Mathematical programming (MP), such as integer programming (IP)
  and mixed integer programming (MIP).
\item Global optimisation (GO).
\item Evolutionary algorithms (EA).
\item Computational geometry.
\item Ad hoc algorithm design.
\item etc.
\end{itemize}
Hybrid optimisation technologies are becoming increasingly powerful,
witness the hybridisation of CP with SLS \cite{Comet}, called
constraint-based local search (CBLS), and the hybridisation of CP with
MP and GO \cite{Hooker:integrated}.

\paragraph{Test Scale.}
An airspace sectorisation tool can be tested at different scales:
\begin{itemize}
\item \emph{Continental}: A (large fraction of an) entire continent is
  sectorised, such as the European Civil Aviation Conference (ECAC)
  area, or its core countries along the London-Frankfurt-Rome axis.
\item \emph{AOS}: An existing area of specialisation is (re-)sectorised.
\item \emph{ATCC}: An (existing) ATCC is (re-)sectorised.
\end{itemize}

\paragraph{Test Data.}
An automated airspace sectorisation tool can be tested on different
kinds of data:
\begin{itemize}
\item \emph{Historical data} stem from archives of actual flight data.
\item \emph{Extrapolated data} is computed from historical data
  according to some forecast of future flight patterns and volumes.
\item \emph{Artificial data} is generated randomly according to some
  (ideally realistic) model of flight patterns and volumes.
\item \emph{Simulated data} results from fast-time simulations of
  historical or extrapolated flight schedules.
\end{itemize}

\section{Survey}
\label{sect:survey}

In the following, we survey the (in our opinion) most important
papers, proceeding in chronological order, breaking ties by alphabetic
order on the surname of the first author.  Unless otherwise indicated,
each constraint is hard and explicit.  A word flagged with a question
mark (`?') means that we think it is correct but found no explicit
confirmation for it in the discussed paper.

\subsection{\cite{Delahaye:ICEC98}}

\begin{center}
\begin{tabular}{|l|l|}
  \hline
  Criterion & Value \\
  \hline\hline
  Approach & graph-based \\ \hline
  Frequency & (any) \\ \hline
  Input granularity & (not applicable) \\ \hline
  Output granularity & (any) \\ \hline
  Dimensionality & 3D \\ \hline
  Constraints & \parbox{11.5cm}{balanced workload, minimum
    dwell time, minimum distance, trajectory-based convexity,
    connectedness} \\ \hline
  Cost function & minimal coordination workload \\ \hline
  Technology & EA: genetic algorithm \\ \hline
  Test scale & ATCC \\ \hline
  Test data & artificial \\ \hline
\end{tabular}
\end{center}
This is one of the oldest and most cited lines of research on
sectorisation.  No quantitative definition is given of the workload of
a sector.  In the post-processing to the graph partitioning, 3D
Vorono\"i diagrams are used to define precisely the 3D borders of
sectors.  Problem instances of up to $400$ vertices that are to be
partitioned into up to $100$ sectors are solved in reasonable time.

\subsection{\cite{Yousefi:ATIO04}}

\begin{center}
\begin{tabular}{|l|l|}
  \hline
  Criterion & Value \\
  \hline\hline
  Approach & region-based \\ \hline
  Frequency & any \\ \hline
  Input granularity & hexagonal mesh, each side of a hexagon
    being $24$ NM \\ \hline
  Output granularity & elementary sectors \\ \hline
  Dimensionality & 2.5D \\ \hline
  Constraints & \parbox{11.5cm}{balanced workload, convexity (soft),
    reasonable average dwell time, connectedness, compactness (soft)} \\ \hline
  Cost function & minimal variation of workload among sectors \\ \hline
  Technology & MIP: facility location problem (number of sectors is
    not fixed) \\ \hline
  Test scale & continental: USA; initial experiments are with an
    ATCC \\ \hline
  Test data & extrapolated: TAAM simulation of one day of traffic \\ \hline
\end{tabular}
\end{center}
Workload is the sum of four components: the horizontal movement
workload, the conflict detection and resolution workload, the
coordination workload, and the altitude-change workload.  The
\emph{horizontal movement workload} is determined by the number of
aircraft in a sector and the average flight time.  The \emph{conflict
  detection and resolution workload} is based on conflict detection
using the type of conflict and the conflict severity.  The
\emph{coordination workload} is determined by the type of coordination
action including voice call, clearance issue, inter-facility transfer,
and tower transfer.  The \emph{altitude-change workload} is determined
by the type of sector altitude clearance requested.  No details are
given on how to compute the workload, but the reader is referred to
\cite{Yousefi:ATM03}, where more details are given, but the
definitions still depend on unspecified adjustment factors.  The tool,
TAAM, used to simulate the traffic produces workload estimates based
on a model of control workload.  The airspace is divided into three
layers: FL0 -- FL210, FL210 -- FL310, plus FL310 and above.
No details are given of the constraints and cost function, nor is
there any definition of what it means to minimise the variation of
workload among the sectors.

\subsection{\cite{TranDac:RAIRO05}}

\begin{center}
\begin{tabular}{|l|l|}
  \hline
  Criterion & Value \\
  \hline\hline
  Approach & graph-based \\ \hline
  Frequency & (any) \\ \hline
  Input granularity & (not applicable) \\ \hline
  Output granularity & (any) \\ \hline
  Dimensionality & 2D \\ \hline
  Constraints & \parbox{11.5cm}{balanced monitoring + conflict
    workload, minimum dwell time, minimum distance,
    trajectory-based convexity, connectedness} \\ \hline
  Cost function & minimal coordination workload \\ \hline
  Technology & hybrid of CP and SLS \\ \hline
  Test scale & ATCC \\ \hline
  Test data & random \\ \hline
\end{tabular}
\end{center}
Building on the model of \cite{Delahaye:ICEC98} (except that the
coordination workload is not considered within the balance
constraint), this work develops a CP model (rather than an EA one),
introducing propagators for new global constraints and introducing new
branching heuristics.  The monitoring and conflict workload of a
sector is the number of aircraft entering the sector.  The
coordination workload of a sector is the number of aircraft leaving
the sector for an adjacent one.  To solve large instances, the problem
is solved with a hybrid algorithm obtaining first a good solution with
an SLS heuristic and then locally improving the solution with an exact
CP formulation on small subsets of the sectors.  Geometrical sectors
are then built with triangulation techniques similar to the ones used
by \cite{Delahaye:ICEC98}.  Problem instances of up to $1000$ vertices
that are to be partitioned into up to $80$ sectors are solved in
reasonable time.

\subsection{\cite{Bichot:ATM07}}

\begin{center}
\begin{tabular}{|l|l|}
  \hline
  Criterion & Value \\
  \hline\hline
  Approach & region-based \\ \hline
  Frequency & static \\ \hline
  Input granularity & elementary sectors \\ \hline
  Output granularity & ATCCs or FABs \\ \hline
  Dimensionality & 3D \\ \hline
  Constraints & \parbox{11.5cm}{balanced monitoring + conflict
    workload (within a factor of $2$), balanced size (within a
    factor of $2$)} \\ \hline
  Cost function & minimal coordination workload \\ \hline
  Technology & SLS: fusion-fission metaheuristic \\ \hline
  Test scale & continental: $11$ core ECAC countries of Europe \\ \hline
  Test data & historical \\ \hline
\end{tabular}
\end{center}
This work\footnote{This paper is classified by \cite{Zelinski:DASC11}
  as work on configuration, but we disagree and include it in our
  survey.} tackles sectorisation at a larger granularity, namely the
design of ATCCs or FABs from the existing elementary sectors (along
mostly national boundaries), so as to enforce regional cooperation in
the seamless management and control of traffic flows in the ECAC zone.
The monitoring and conflict workload of a sector is the daily number
of aircraft entering the sector.  The coordination workload of a
sector is the daily number of aircraft leaving the sector for an
adjacent one.  A novel meta-heuristic, called \emph{fusion-fission}
and inspired by the corresponding phenomena in nuclear physics (but
described in detail only in a French-medium journal), is shown to be
particularly well-suited to solve this problem.  It seems to
outperform consistently very powerful graph partitioning libraries and
outputs FABs that much improve on the coordination workload and
imbalance of the current ATCC partition of ECAC.

\subsection{\cite{Conker:ATM07}}

\begin{center}
\begin{tabular}{|l|l|}
  \hline
  Criterion & Value \\
  \hline\hline
  Approach & region-based \\ \hline
  Frequency & static \\ \hline
  Input granularity & square mesh, dimension not defined \\ \hline
  Output granularity & elementary sectors \\ \hline
  Dimensionality & 2D(?) \\ \hline
  Constraints & equal complexity density across sectors (soft) \\ \hline
  Cost function & (none) \\ \hline
  Technology & ad hoc algorithm: $k$-means clustering, with an SLS heuristic \\ \hline
  Test scale & continental: USA, west of the Mississippi \\ \hline
  Test data & \parbox{11.5cm}{extrapolated: by simulation of traffic
    patterns based on predicted traffic demands} \\ \hline
\end{tabular}
\end{center}
A combination of two existing tools
\cite{bhadra2005future,wanke2004progressive} and specialised
algorithms is proposed.  The aim is to give a tool chain that assists
the semi-automated design of new sectors.  The workload is defined in
terms of a complexity density metric that is not defined in any
detail.  For the full definition of complexity density the reader is
referred to an unpublished MITRE Corporation memo.  The algorithmic
phase has three components: a modification of $k$-means clustering
\cite{lloyd:k-means:1982} to produce an initial sectorisation; a
hand-crafted SLS heuristic to improve the workload balance; and
straightening of the sector boundaries.  Many of the technical details
of the algorithms are missing.  Much of the effort is dedicated to
tool support to allow human experts to assess the given sectorisation.
Two tools are described: \emph{airspaceAnalyzer}, which simulates air
traffic controllers to assess the workload balance of the new
sectorisation; and \emph{sectorEvaluator}, which allows experts to
assess the quality of a given sectorisation and modify its sectors.

\subsection{\cite{Martinez:GNC07}}

\begin{center}
\begin{tabular}{|l|l|}
  \hline
  Criterion & Value \\
  \hline\hline
  Approach & combination of graph-based and region-based \\ \hline
  Frequency & static \\ \hline
  Input granularity & (not applicable) \\ \hline
  Output granularity & control sectors \\ \hline
  Dimensionality & 2D \\ \hline
  Constraints & bounded workload \\ \hline
  Cost function & minimal number of sectors \\ \hline
  Technology & ad hoc algorithm based on spectral clustering \\ \hline
  Test scale & continental: USA \\ \hline
  Test data & historical: ASDI (aircraft situation display to industry) data \\ \hline
\end{tabular}
\end{center}
Definitions of workload are not considered in any detail.  Instead
workload balancing is done in terms of the peak traffic count in a
sector based on an unspecified time interval.  The constraints and the
cost functions are defined by ad hoc algorithms.  The algorithm is in
three stages: first, flight data is used to construct a network of
flight flows that correspond to frequently flown routes; then spatial
cells are assigned to nodes of the flow network based on a nearest
neighbour, where each node in the flow is weighted by the number of
flights passing through that node; finally, the flow network is
partitioned into smaller and smaller sub-graphs until all sub-graphs
have a workload cost below a certain threshold.  Since the
partitioning problem is NP-hard, a heuristic based on spectral
techniques \cite{simon1991partitioning} from graph theory is used.

\subsection{\cite{Brinton:ATIO09}}

\begin{center}
\begin{tabular}{|l|l|}
  \hline
  Criterion & Value \\
  \hline\hline
  Approach & region-based \\ \hline
  Frequency & static \\ \hline
  Input granularity & square mesh, each side of a square being 1NM \\ \hline
  Output granularity & elementary sectors \\ \hline
  Dimensionality & 2D \\ \hline
  Constraints & non-jagged boundaries (soft), bounded workload \\ \hline
  Cost function & minimal number of sectors \\ \hline
  Technology & hybrid: computational geometry and SLS \\ \hline
  Test scale & continental USA \\ \hline
  Test data & not specified \\ \hline
\end{tabular}
\end{center}
The definition of workload is based on dynamic density, of which there
are multiple definitions in \cite{Kopardekar:DASC02}, but the one used
in this study is not specified.  The algorithm has three stages: an
initial clustering of flight tracks to produce an initial air-space
partition; a grid-based approach, where grid cells can have zero, one,
or many clusters, is used to grow cells containing clusters to
candidate sectors; and finally an SLS algorithm is used to simplify or
straighten the sector boundaries and to combine candidate sectors
while keeping the peak dynamic density below a certain limit.

\subsection{\cite{Drew:ATIO09}}

\begin{center}
\begin{tabular}{|l|l|}
  \hline
  Criterion & Value \\
  \hline\hline
  Approach & region-based \\ \hline
  Frequency & dynamic \\ \hline
  Input granularity & elementary sectors: free-form \\ \hline
  Output granularity & control sectors \\ \hline
  Dimensionality & 3D \\ \hline
  Constraints & bounded workload, connectedness \\ \hline
  Cost function & minimal number of sectors \\ \hline
  Technology & MIP \\ \hline
  Test scale & ATCC: Cleveland high sectors \\ \hline
  Test data & historical \\ \hline
\end{tabular}
\end{center}
The workload of a sector is the maximum number of aircraft
simultaneously present in the sector over a $15$-minute interval; it
has a sector-specific upper bound called the monitor alert parameter
(MAP).  The absence of a shape constraint, such as compactness, is
noted to lead to convoluted control sectors.  The presence of
symmetries in the solution space is noted, but no way is proposed to
exploit them.

\subsection{\cite{Leiden:ATIO09}}

\begin{center}
\begin{tabular}{|l|l|l|}
  \hline
  Criterion & \multicolumn{2}{l|}{Value} \\
  \hline\hline
  Approach & \multicolumn{2}{l|}{region-based} \\ \hline
  Frequency & \multicolumn{2}{l|}{dynamic: every $15$ minutes to $24$ hours} \\ \hline
  Input granularity & \multicolumn{2}{l|}{one AOS or ATCC: base-line} \\ \hline
  Output granularity & \multicolumn{2}{l|}{control sectors} \\ \hline
  Dimensionality & \multicolumn{2}{l|}{3rd D} \\ \hline
  Constraints (shared) & \multicolumn{2}{l|}{connectedness,
    lower-bounded height ($2$ FL)} \\ \hline
  Constraints (specific) & \parbox{5.1cm}{balanced workload \\
    (fixed number of sectors)}
  & \parbox{5.75cm}{(none other)} \\ \hline
  Cost function & \parbox{5.1cm}{(none)}
  & \parbox{5.75cm}{minimal number of sectors} \\ \hline
  Technology & \multicolumn{2}{l|}{ad hoc algorithm design: greedy algorithm} \\ \hline
  Test scale & \multicolumn{2}{l|}{AOS or ATCC: USA, above FL$240$} \\ \hline
  Test data & \multicolumn{2}{l|}{historical} \\ \hline
\end{tabular}
\end{center}
The workload of a sector is the number of aircraft simultaneously
present in the sector over an unspecified time interval; it has a
sector-independent upper bound called the monitor alert parameter
(MAP).  If workload exceeds the MAP, then a horizontal splitting of
the sector must be performed, but how to do this has not been
investigated yet.  The algorithm works in two modes, as indicated in
the table above.  Either way, the algorithm picks the best solution
from a greedy bottom-up phase and a greedy top-down phase through the
considered airspace.  The transition cost between two sectorisations
is the number of aircraft that must be transferred into a new sector;
this metric is used for evaluating computed sectorisations, but
currently not while computing them.  The experimental evaluation
confirms the expectation that capacity increases with the frequency of
re-sectorisation, but also concludes that the transition cost must be
built into the constraints or cost function.

\subsection{\cite{Li:ATIO09}}

\begin{center}
\begin{tabular}{|l|l|}
  \hline
  Criterion & Value \\
  \hline\hline
  Approach & graph-based \\ \hline
  Frequency & static \\ \hline
  Input granularity & (not applicable) \\ \hline
  Output granularity & control sectors \\ \hline
  Dimensionality & 3D \\ \hline
  Constraints & \parbox{11.5cm}{balanced workload, bounded
    workload, minimum distance, connectedness, compactness} \\ \hline
  Cost function & (none) \\ \hline
  Technology & \parbox{11.5cm}{ad hoc algorithm: spectral clustering,
    convex hull, shortest path (number of sectors is not fixed)} \\ \hline
  Test scale & ATCC: Atlanta \\ \hline
  Test data & historical \\ \hline
\end{tabular}
\end{center}
The workload of a sector is a weighted sum of its monitoring workload
(presumably including the conflict workload) and coordination
workload, though it is not made clear how these terms are quantified.
Special care is taken in the geometric post-processing step to obtain
sectors with smooth boundaries and good shapes.  Excellent results are
obtained, in the sense that the current sectorisation of Atlanta is
outperformed on almost all metrics in both the average case and worst
case.

\subsection{\cite{Bloem:ICAS10}}

\begin{center}
\begin{tabular}{|l|l|}
  \hline
  Criterion & Value \\
  \hline\hline
  Approach & region-based \\ \hline
  Frequency & dynamic \\ \hline
  Input granularity & elementary sectors: base-line \\ \hline
  Output granularity & control sectors \\ \hline
  Dimensionality & 3D \\ \hline
  Constraints & bounded workload (soft), connectedness,
    convexity (soft) \\ \hline
  Cost function & minimal workload cost + transition cost \\ \hline
  Technology & approximate dynamic programming \\ \hline
  Test scale & AOS and ATCC: Cleveland super-high sectors \\ \hline
  Test data & historical \\ \hline
\end{tabular}
\end{center}
The workload of a sector is the maximum number of aircraft in the
sector during the considered time step divided by a sector-specific
upper bound called the monitor alert parameter (MAP).  The transition
cost between two sectorisations is the number of new control sectors
compared to the previous time step.  The uncertainty of trajectory
prediction is explicitly taken into account.  Excellent results are
obtained, in the sense that the current sectorisation of Cleveland is
outperformed.

\subsection{\cite{Gianazza:AI10}}

\begin{center}
\begin{tabular}{|l|l|}
  \hline
  Criterion & Value \\
  \hline\hline
  Approach & region-based \\ \hline
  Frequency & any \\ \hline
  Input granularity & elementary sectors \\ \hline
  Output granularity & control sectors \\ \hline
  Dimensionality & (not applicable) \\ \hline
  Constraints & connectedness \\ \hline
  Cost function & \parbox{11.5cm}{minimal number of sectors + 
    workload predictions obtained from the neural network} \\ \hline
  Technology & complete search via branch-and-bound \\ \hline
  Test scale & ATCC: five French ATCCs \\ \hline
  Test data & historical data \\ \hline
\end{tabular}
\end{center}
The workload of a sector is defined using a neural network trained on
historical traffic data and sectorisations.  The work considers
combinations of elementary airspace modules to build an optimal
airspace partition.  The work differs from other papers in this survey
in that it uses a neural network to predict workloads of
configurations and tree search together with branch-and-bound
techniques to explore intelligently all possible combinations of
elementary sectors in order to find the optimal combination.

\subsection{\cite{Sabhnani:ATIO10}}

\begin{center}
\begin{tabular}{|l|l|}
  \hline
  Criterion & Value \\
  \hline\hline
  Approach & region-based and graph-based \\ \hline
  Frequency & dynamic \\ \hline
  Input granularity & elementary sectors: free-form \\ \hline
  Output granularity & control sectors \\ \hline
  Dimensionality & 2D with extensions to 2.5D \\ \hline
  Constraints & \parbox{11.5cm}{balanced monitoring + conflict
    workload, compactness, minimum distance, convexity, and two other
    operational constraints (see text below)} \\ \hline
  Cost function & minimal coordination workload \\ \hline
  Technology & hybrid: computational geometry, graph theory, and MIP \\ \hline
  Test scale & ATCC \\ \hline
  Test data & simulated data \\ \hline
\end{tabular}
\end{center}
The focus is to extend previous work \cite{Mitchell:GNC08} to build
sectors that satisfy controllers (operational criteria).  It is argued
that while workload balance is an important goal, the validity of the
sector shapes is also important.  The major operational criteria that
are addressed are that: standard flows should cross sector boundaries
(almost) orthogonally; critical merge-and-conflict pairs should remain
sufficiently inside sector boundaries (minimum distance); no more than
three sectors should come together at the same point; and the shape of
the sector show be more or less convex.  The definition of workload is
as in \cite{Mitchell:GNC08}, where three possible metrics are
considered: \emph{peak workload}, which is the maximum number of
aircraft simultaneously in a sector; \emph{average workload}, which
the average number of aircraft present in a sector; and
\emph{coordination workload}, which is the number of instances during
an unspecified time interval where an aircraft crosses the boundary of
a sector.

\subsection{\cite{Xue:GNC10}}

\begin{center}
\begin{tabular}{|l|l|}
  \hline
  Criterion & Value \\
  \hline\hline
  Approach & region-based \\ \hline
  Frequency & static \\ \hline
  Input granularity & elementary sectors: free-form \\ \hline
  Output granularity & control sectors \\ \hline
  Dimensionality & 3D \\ \hline
  Constraints & (all in the cost function) \\ \hline
  Cost function & \parbox{11.5cm}{maximal average dwell time; maximal distance of
    intersection points from sector boundaries; maximal distance of
    dominant flows from sector boundaries; minimal number of
    flights with short dwell time; minimal variance of sector peak
    flight counts} \\ \hline
  Technology & hybrid: computational geometry and EA: genetic algorithm \\ \hline
  Test scale & ATCC \\ \hline
  Test data & simulated data \\ \hline
\end{tabular}
\end{center}
There is no explicit definition of workload.  The algorithms use a
mixture of computational geometry (Vorono\"i diagrams) and genetic
algorithms with a post-processing step using iterative deepening
search to improve the resulting sector designs.

\subsection{\cite{Jaegare:MSc11}}

\begin{center}
\begin{tabular}{|l|l|}
  \hline
  Criterion & Value \\
  \hline\hline
  Approach & region-based \\ \hline
  Frequency & (any) \\ \hline
  Input granularity & hexagonal or square mesh (arbitrary size) and AFBs \\ \hline
  Output granularity & elementary sectors \\ \hline
  Dimensionality & 3D \\ \hline
  Constraints & \parbox{11.5cm}{balanced monitoring + conflict
    workload, minimum dwell time, minimum distance,
    trajectory-based convexity} \\ \hline
  Cost function & minimal number of entry points \\ \hline
  Technology & CP \\ \hline
  Test scale & ATCC: Europe \\ \hline
  Test data & extrapolated, by ASTAAC \\ \hline
\end{tabular}
\end{center}
This work is close in spirit to \cite{TranDac:RAIRO05}, but takes a
region-based approach (rather than a graph-based one).  It also takes a
CP approach, but without hybridisation, and also introduces
propagators for new global constraints.
The monitoring and conflict workload of a sector is the number of
aircraft entering the sector.  The coordination workload of a sector
is the number of aircraft leaving the sector for an adjacent one.
The data for the experiments is provided by the \emph{Arithmetic
  Simulation Tool for ATFCM and Advanced Concept} (ASTAAC) tool of the
\EuroControl\ Experimental Centre.  This tool actually pre-clusters
some regions into AFBs according to additional constraints (sufficient
distance from potential conflicts and trajectories to sector
boundaries).
Experiments were run on up to a few hundred thousand small regions, to
be partitioned into five sectors.  In comparable run-times on the same
machine, the constraint program produces much better sectorisations
than \emph{NEVAC Sector Builder}, which is a greedy algorithm that
comes with ASTAAC (but is unpublished) and considers the
trajectory-based-convexity and minimum-dwell-time constraints to be
soft and yet does not systematically yield fewer sector entry points.
Since two hard constraints, connectedness and compactness, were not
implemented yet, the resulting sectors are sometimes disconnected or
of highly irregular shapes, so that current air traffic controllers
would be very uncomfortable in working with them.
The work is currently being revisited with us (who supervised this
thesis work), with the following targeted profile:
\begin{center}
\begin{tabular}{|l|l|}
  \hline
  Criterion & Value \\
  \hline\hline
  Approach & region-based \\ \hline
  Frequency & (any) \\ \hline
  Input granularity & hexagonal or square mesh (arbitrary size) and AFBs \\ \hline
  Output granularity & elementary sectors \\ \hline
  Dimensionality & 3D \\ \hline
  Constraints & \parbox{11.5cm}{balanced monitoring + conflict
    workload (soft), minimum dwell time (soft), trajectory-based
    convexity (soft), compactness (soft), connectedness} \\ \hline
  Cost function & minimal violation of soft constraints \\ \hline
  Technology & SLS: CBLS \\ \hline
  Test scale & continental: Europe \\ \hline
  Test data & extrapolated, by ASTAAC \\ \hline
\end{tabular}
\end{center}

\subsection{\cite{Kulkarni:ICNS11}}

\begin{center}
\begin{tabular}{|l|l|}
  \hline
  Criterion & Value \\
  \hline\hline
  Approach & region-based \\ \hline
  Frequency & static \\ \hline
  Input granularity & hexagonal mesh, each side of a hexagon being $24$ NM \\ \hline
  Output granularity & elementary sectors \\ \hline
  Dimensionality & 3D \\ \hline
  Constraints & balanced workload, compactness(?) \\ \hline
  Cost function & (none) \\ \hline
  Technology & ad hoc algorithms: knapsack, set covering, etc. \\ \hline
  Test scale & continental: USA, above FL240 \\ \hline
  Test data & extrapolated, by PNP \\ \hline
\end{tabular}
\end{center}
The paper completely lacks definitions and algorithmic details.  It is
claimed (partly in the title) that approximate dynamic programming as
well as control theory are involved, but it is not apparent where.

\section{Conclusion}
\label{sect:concl}

Sectorisation would benefit from further modelling in order to adapt
better to the design of FABs, taking into account more realistic
operational constraints, such as compatibility constraints between
upper and lower airspace, and those of \cite{Sabhnani:ATIO10} and
\cite{Xue:GNC10}.  The implementation of airspace optimisations that
alter the shape of basic control sectors implies a heavy cost in
controller training, as controllers are highly specialised in the
management of their specific sectors and it takes several years to
qualify them on new sectors.  Moreover, the redesign of ATCCs would
induce a new dispatch of radar data and probably the building of
costly new infrastructures to host them.  Contrary to lighter
optimisations that only concern current control structures without
modifying them, like the optimisation of opening schemes changes in
airspace design have very important transition costs and must be very
carefully planned.

From an algorithmic viewpoint, a clear conclusion of this survey is
that constraints should more often be used in the \emph{process} of
computing a sectorisation, rather than only in evaluating the
\emph{results} of a sectorisation algorithm.  We argue that mature
optimisation technologies, such as CP and MP, should be used more
often.  On the one hand, they offer high-level modelling facilities,
so that one can make \emph{explicit} the constraints and cost
function.  On the other hand, they offer off-the-shelf solving
algorithms that operate directly on such models.  In other words,
there is a clean separation of concerns between the declarative aspect
of modelling and the procedural aspect of solving, and highly tuned
algorithms can be re-used.  Such a plug-and-play paradigm allows a
much nimbler prototyping exploration of the algorithm design space,
which is crucial in an area such as sectorisation where suitable sets
of constraints and cost functions still have to be identified.

\bibliographystyle{apalike}
\bibliography{sectorisation}

\end{document}